# Exchange bias up to room temperature in the antiferromagnetic bulk hexagonal $Mn_3Ge$

J. F. Qian,[1,a] A. K. Nayak,[1] G. Kreiner,[1] W. Schnelle,[1] and C. Felser[1,b]

*Max Planck Institute for Chemical Physics of Solids, Nöthnitzer Str. 40, 01187 Dresden, Germany*

(Dated: 12 November 2013)

This work reports an exchange bias (EB) effect up to room temperature in the binary intermetallic bulk compound $Mn_{3.04}Ge_{0.96}$. The sample annealed at 700 K crystallizes in a tetragonal structure with ferromagnetic ordering, whereas, the sample annealed at 1073 K crystallizes in a hexagonal structure with antiferromagnetic ordering. The hexagonal $Mn_{3.04}Ge_{0.96}$ sample exhibits an EB of around 70 mT at 2 K that continues with a non-zero value up to room temperature. The exchange anisotropy is proposed to be originating from the exchange interaction between the triangular antiferromagnetic host and the embedded ferrimagnetic like clusters. The ferrimagnetic clusters develop when excess Mn atoms occupy empty Ge sites in the original triangular antiferromagnet structure of $Mn_3Ge$.

Binary Heusler compounds Mn3Z, containing man-ganese with a main group element Z, exhibit various structural and magnetic properties.[1-4] In this family, $Mn_3Ga$ has attracted a lot of attention due to its possible application in spin transfer torque (STT) device and hard magnetic behavior.[5,6] In addition, doped with a second transition element the structure and magnetic properties can be tuned to fit for various applications. For example, the Co substitution in place of Mn in $Mn_{3-x}Co_xGa$ transfers the tetragonal distorted Heusler structure known as a hard magnet to a cubic soft magnet with half-metallic behavior.[7] Similarly, $Mn_2PtGa$ has been found to remain in the tetragonal phase and exhibits a large zero field cooled exchange bias (EB).[8] The tetragonal compounds are interesting as they show perpendicular magnetic anisotropy in thin films which opens a way for application in STT magnetic random access memories (MRAM).[9-12] Moreover, a hexagonal triangular antiferromagnetic film of hexagonal-Mn3Ga has been used as a pinned layer in magnetic tunnel junctions (MTJ).[13] Besides the high tunneling magneto-resistance (TMR), a large EB field also has been observed in this system.[13] They attributed this EB to the frustrated triangular antiferromagnetic structure easily perturbed by exchange coupling to the adjacent ferromagnetic layer.

In the present work, we focus on another binary compound Mn3Ge, which was first synthesized as a hexagonal structure in 1949 by Zwicker et al.[14] An antiferromagnetic (AFM) ordering with feeble parasitic ferromagnetic (FM) behavior was observed in

[a] Jin-Feng.Qian@cpfs.mpg.de
[b] felser@cpfs.mpg.de

Mn$_{3.4}$Ge.[4] A neutron diffraction study in Mn$_{3.1}$Ge$_{0.9}$ have revealed a triangular AFM structure with magnetic moment lying in the hexagonal basal plane.[15] A Néel temperature $T_N$ = 395 ±10 K has been reported for this compound.[15] The small FM component was attributed to the free rotation of the spin triangle in the basal-plane caused by a small hexagonal distortion.[16] The hexagonal structure transforms to a face-centered tetragonal phase by a long annealing at 450 °C for more than ten days.[4] A large perpendicular magnetic anisotropy and TMR have been reported in the tetragonal Mn$_3$Ge thin film.[17] However, there is not enough study on the magnetic properties of the hexagonal Mn$_3$Ge. In this letter we present a detailed structural and magnetic study on the bulk Mn$_3$Ge. We show that the hexagonal-Mn$_3$Ge sample exhibits exchange bias up to room temperature.

Polycrystalline ingots of Mn$_3$Ge were carefully synthesized by repeated inductive melting the target amounts of high purity elements in a high purity argon atmosphere under 10$^{-2}$ mbar pressure in a glove box. During melting, the metal pieces were contained in a cleaned alumina crucible. In consideration of the high vapor pressure of Mn, and avoiding Mn loss, the induction current was controlled under an appropriate value. The weight loss after the whole process was less than 1%. Afterward, the samples were sealed in evacuated quartz tubes to anneal at 1073 K for 4 days. In order to get higher degree of atomic ordering a part of the sample was further annealed at 823 K for 10 days and quenched into a mixture of ice and water. For comparison, another part of the sample was annealed at 700 K for 14 days to obtain the low temperature tetragonal-Mn3Ge phase. The sample quality and composition were studied by utilizing a scanning electron microscope equipped with an energy dispersive x-ray spectroscopy (EDXS) detection system and inductively coupled plasma optical emission spectrometry (ICP-OES). The crystal structure was investigated by using powder x-ray diffraction (XRD), with Cu-Kα ($\lambda$ = 1.540598 Å) radiation. The magnetic properties were studied using a superconducting quantum interference device (SQUID, Quantum Design MPMS) magnetometer between 1.8 K and 400 K.

From the EDXS and ICP-OES analysis it is found that the hexagonal Mn3Ge consists of a composition of 76.58:23.42, whereas, the tetragonal Mn3Ge sample has a composition of 76.62:23.38. In both cases the compositions are very close to the initial composition of 75:25. The phase purity of both the samples is confirmed by theoretical fitting to the experimental XRD data as shown in Fig. 1. In case of 1073 K annealed sample (Fig. 1a) the refinement has been performed with a hexagonal space group P63/mmc, which gives the lattice parameters a = 5:34 Å and c = 4:32 Å. The structure matches to the typical D0$_{19}$ hexagonal unit cell as expected. Similarly, for the 700 K annealed sample (Fig. 1b) the refinement has been performed with a tetragonal space group I4/mmm, which results in lattice parameters a = 3:80 Å and c = 7:24 Å. From the fitting it is clear that both samples don't show reflexes of secondary phases or impurities. The hexagonal Mn3Ge has three pairs of Mn atoms and one pair of Ge atoms for unit cell (inset of Fig. 1a). The basic magnetic structure of Mn atoms is inverse triangular, which means the chirality of the two spins triangles in a unit cell are opposite.

The temperature variation of the magnetization, *M(T)*, for Mn$_3$Ge measured in zero field cooled (ZFC) and field cooled (FC) modes are shown in Fig. 2. In the ZFC mode, the sample was initially cooled from 400 K down to 2 K in zero field and data were taken upon increasing temperature in applied field, whereas, in the FC mode, data were collected while cooling in field. The ZFC and FC *M(T)* measured at various fields for hexagonal Mn3Ge are shown in Fig. 2a. The sample undergoes a magnetic order disorder transition around 370 K. This transition was previously recognized as the Curie temperature T$_C$ connected with the residual ferromagnetic (FM) component of the AFM structure,[18] which is very close to the Néel temperature ($T_N$ = 395 ±10 K) of its antiferromagnetic structure.15 The residual FM component arises from the parent AFM phase as a result of small distortion of the hexagonal structure giving rise to a tiny uncompensated FM component.[15] The ZFC *M(T)* curve measured in 0.1 T follows a lower magnetization path than the FC curve, which makes a large irreversibility between the two curves. This irreversibility continues for the *M(T)* measured up to 1 T. This indicates the presence of a magnetic inhomogeneous state in the whole temperature regime though the sample stays homogeneous structurally. The deviation between ZFC and FC diminishes with increasing the measuring field but does not disappear completely up to 7 T (Fig. 2b). To compare the magnetic state of the tetragonal Mn$_3$Ge, we have measured the high temperature *M(T)* in a field of 1 T as shown in Fig. 2c. The sample shows a TC of around 800 K, which is interesting for application prospective.

The magnetic hysteresis loops measured at 2 K after zero field cooling the samples from 400 K for both the hexagonal and the tetragonal phases are shown in Fig. 3. As expected the hexagonal sample shows a very small magnetic moment of around 0.09$\mu_B$/f.u. at 7 T as it arises from the little uncompensated moment of the AFM structure. In addition, the linear and unsaturated magnetic behavior at higher fields clearly indicates the AFM nature of the sample. Due to the strong magneto-crystalline anisotropy of the hexagonal structure, the sample shows a large coercive field, $H_C$, of around 0.5 T even though a small FM component present in the structure. The neutron diffraction study in the hexagonal-Mn$_3$Ge suggests $M_S$ =0.02 $\mu_B$/Mn.[15] A little higher magnetic moment in the present case indicates that an additional FM component present in the sample. The tetragonal sample shows a much larger magnetization of 0.7$\mu_B$/f.u. at 7 T. This indicates a normal ferrimagnetic ordering as observed in other Mn rich compounds.[19,20] In tetragonal phase the sample shows a $H_C$ of only 0.27 T.

The presence of magnetic irreversibility in the hexagonal Mn$_3$Ge (Fig. 2) indicates that there might be some extra FM component in the nearly AFM background. Therefore we have measured field cooled hysteresis loops to map out any possible interaction between the AFM host and FM clusters. The field cooled hysteresis loops measured at different temperatures are shown in Fig. 4a. In the field cooled mode a field of 5 T was applied at 400 K before cooling the sample to measurement temperature. All measurements were performed in ±7 T range. It is found that all loops are shifted in negative field direction indicating the presence of exchange bias (EB) in the sample. The temperature dependence of the EB field ($H_{EB}$) and the coercive field ($H_C$) calculated from the FC hysteresis loops are shown in Fig. 4b. $H_{EB}$ and $H_C$ are calculated using $H_{EB} = |H_1+H_2|/2$ and $H_C = |H_1-H_2|/2$, where $H_1$ and $H_2$ are the lower and upper cut-off fields. A maximum $H_{EB}$ =70 mT has been calculated at 2 K. The EB monotonically decreases with temperature that holds a small non-zero value at room temperature. Similarly, $H_C$ of 0.47 T and 0.25 T have been calculated at 2 K and 300 K, respectively. The cooling field, $H_{CF}$, dependence of EB and coercive field are plotted in the inset of Fig. 4b. Both $H_{EB}$ and $H_C$ increase with cooling field up to 5 T that show saturation behavior for higher fields. The tetragonal Mn3Ge sample does not show any detectable exchange bias behavior. It has to be mentioned here that though EB has been observed in various bulk materials including several Heusler alloys, the effect in most of the cases has been limited to the low temperatures.[21-27] The present finding of EB up to room temperature certainly is a subject of interest for application point of views.

Now that exchange bias has been established in the hexagonal Mn$_3$Ge sample, it is clear that there must exist two different magnetically anisotropic states in the sample, one being the AFM host, the other must be with a higher magnetic state. As mentioned earlier the EDXS and ICP-OES analysis give an average composition of 3.04:0.96 Mn to Ge ratio. This implies that for some unit cells the extra Mn atoms occupy the Ge sites. As shown in Fig. 1a the magnetic moments of three Mn atoms lying in the basal plane constitute a triangular antiferromagnetic structure. In case of Mn replacing Ge atom in the same plane, it will feel a strong magnetic frustration from the three basal plane Mn atoms and the other three in the interlayer. In this process the overall magnetic moment will increase to give a ferrimangetic like situation to decrease the frustration. Since the distances between the Mn atoms are not equal, a ferromagnetic-like structure is feasible. The present scenario will result in some ferrimagnetic clusters in the host AFM background. Hence, the exchange interaction between the AFM host and the ferrimagnetic clusters is proposed to rise to the observed exchange anisotropy.

In conclusion, we present the structural and magnetic properties of bulk Mn$_3$Ge (with excess Mn atomes). With different annealing conditions one can obtain a hexagonal and a tetragonal structure that display antiferromagnetic and ferrimagnetic ordering, respectively. We demonstrate the presence of exchange bias (EB) up to room temperature in the hexagonal sample. The EB in the present case arises from the exchange interaction between the triangular AFM host and the ferrimagnetic clusters. The excess Mn atoms occupying the empty Ge sites convert the original triangular antiferromagnet to a ferrimagnet without changing the crystal structure. The existence of EB up to room temperature in the bulk hexagonal $Mn_3Ge$ is undoubtedly promising for potential application as most of the bulk materials show EB only below room temperature. As the EB possesses a great importance in memory devices the present work will encourage further studies in thin films for application purpose.

This work was financially supported by the Deutsche Forschungsgemeinschaft DFG (Project No.TP 2.3-A of Research Unit FOR 1464 ASPIMATT) and by the ERC Advanced Grant (291472) "Idea Heusler".

[1]E. Kren, J. Paitz, G. Zimmer and E. Zsoldos, J Physica B 80,226 (1975).

[2]S. Tomiyoshi, Y. Yamaguchi, M. Ohashi, E. R. Cowley and G. Shirane, Phys. Rev. Lett. 36, 2181 (1987).

[3]E. Kren and G. Kadar, Solid State Commun. 8, 1653 (1970).

[4]T. Ohyoyama, J. Phys. Soc. Japan 16, 1995 (1961).

[5]B. Balke, G. H. Fecher, J. Winterlik and C. Felser, Appl. Phys. Lett. 90, 152504 (2007).

[6]J. Winterlik, B. Balke, G. H. Fecher, C. Felser, M. C. M. Alves, F. Bernardi, and J. Morais, Phys. Rev. B 77, 054406 (2008).

[7]V. Alijani, J. Winterlik, G. H. Fecher and C. Felser, Appl. Phys. Lett. 99, 222510 (2011).

[8]A. K. Nayak, M. Nicklas, S. Chadov, C. Shekhar, Y. Skourski, J. Winterlik and C. Felser, Phys. Rev. Lett. 110, 127204 (2013).

[9]F. Wu, S. Mizukami, D. Watanabe, H. Naganuma, M. Oogane, Y. Ando and T. Miyazaki, Appl. Phys. Lett. 94, 122503 (2009).

[10]S. Mizukami, F. Wu, A. Sakuma, J. Walowski, D. Watanabe, T. Kubota, X. Zhang, H. Naganuma, M. Oogane, Y. Ando, and T. Miyazaki, Phys. Rev. Lett. 106, 117201 (2011).

[11]H. Kurt, K. Rode, M. Venkatesan, P. Stamenov, and J. M. D. Coey, Phys. Rev. B 83, 020405(R) (2011).

[12]M. Li, X. Jiang, M. G. Samant, C. Felser, and S. S. P. Parkin, Appl. Phys. Lett. 103, 032410 (2013).

[13]H. Kurt, K. Rode, H. Tokuc, P. Stamenov, M. Venkatesan and J. M. D. Coey, Appl. Phys. Lett. 101, 232402 (2012).

[14]U. Zwicker, E. Jahn, and K. Schubert, Z. Metallkunde 40, 433 (1949).

[15]G. Kadar and E. Kren, In. J. Magnetism 1, 143 (1971).

[16]S. Tomiyoshi, Y. Yamaguchi, and T. Nagamiya, J. Magn. Magn. Mater. 31-4, 629 (1983).

[17]H. Kurt, N. Baadji, K. Rode and M. Venkatesan, P. Stamenov,

S. Sanvito and J. M. D. Coey, Appl. Phys. Lett. 101, 132410 (2012).

[18]N. Yamada, H. Sakai, H. Mori, and T. Ohoyama, Physica B C 149, 311 (1988).

[19]J. Winterlik, G. H. Fecher, B. Balke, T. Graf, V. Alijani, V. Ksenofontov, C. A. Jenkins, O. Meshcheriakova, C. Felser, G. Liu, S. Ueda, K. Kobayashi, T. Nakamura, and M. W_ojcik, Phys. Rev. B 83, 174448 (2011).

[20]P. Klaer, C. A. Jenkins, V. Alijani, J. Winterlik, B. Balke, C. Felser, and H. J. Elmers, Appl. Phys. Lett. 98, 212510 (2011).

[21]M. Khan, I. Dubenko, S. Stadler, and N. Ali, Appl. Phys. Lett. 91, 072510 (2007).

[22]A. K. Nayak, K. G. Suresh, and A. K. Nigam, J. Phys. D: Appl. Phys. 42, 115004 (2009).

[23]H. C. Xuan, Q. Q. Cao, C. L. Zhang, S. C. Ma, S. Y. Chen, D. H. Wang, and Y. W. Du, Appl. Phys. Lett. 96, 202502 (2010).

[24]X. D. Tang, W. H. Wang, W. Zhu, E. K. Liu, G. H. Wu, F. B. Meng, H. Y. Liu, and H. Z. Luo, Appl. Phys. Lett. 97 (2010).

[25]A. K. Nayak, R. Sahoo, K. G. Suresh, A. K. Nigam, X. Chen, and R. V. Ramanujan, Appl. Phys. Lett. 98, 232502 (2011).

[26]A. K. Nayak, C. Shekhar, J. Winterlik, A. Gupta, and C. Felser, Appl. Phys. Lett. 100, 152404 (2012).

[27]R. Machavarapu and G. Jakob, Appl. Phys. Lett. 102, 232406 (2013).

Figure Captions

FIG. 1. (Color online) Room temperature powder x-ray diffraction (XRD) of Mn3Ge (a) annealed at 1073 K and (b) annealed at 700 K. The black lines show the experimental data and the red lines correspond to the fitting. The inset of (a) shows the crystal structure of the hexagonal phase, whereas, the inset of (b) shows the crystal structure of tetragonal phase. Mn atoms (blue spheres), Ge atoms (green spheres).

FIG. 2. (Color online) Temperature dependence of magnetization $M(T)$ for the hexagonal and tetragonal Mn3Ge. (a) Zero field cooled (ZFC) and field cooled (FC) $M(T)$ of hexagonal Mn3Ge measured in 0.1 T, 0.5 T and 1 T. (b) ZFC and FC $M(T)$ of hexagonal Mn3Ge measured in 7 T. (c) High temperature $M(T)$ for tetragonal Mn3Ge.

FIG. 3. (Color online) Magnetic isotherms $M(H)$ measured at 2 K for (a) hexagonal Mn3Ge and (b) tetragonal Mn3Ge.

FIG. 4. (Color online) (a) $M(H)$ loops for hexagonal-Mn$_3$Ge measured at different temperatures after field cooling the sample in 5 T. (b) Temperature dependence of exchange bias field $H_{EB}$ and coercive field $H_C$ calculated from the field cooled $M(H)$ loops. Inset of (b) shows the cooling field, $H_{CF}$, dependence of $H_{EB}$ and $H_C$ measured at 2 K.

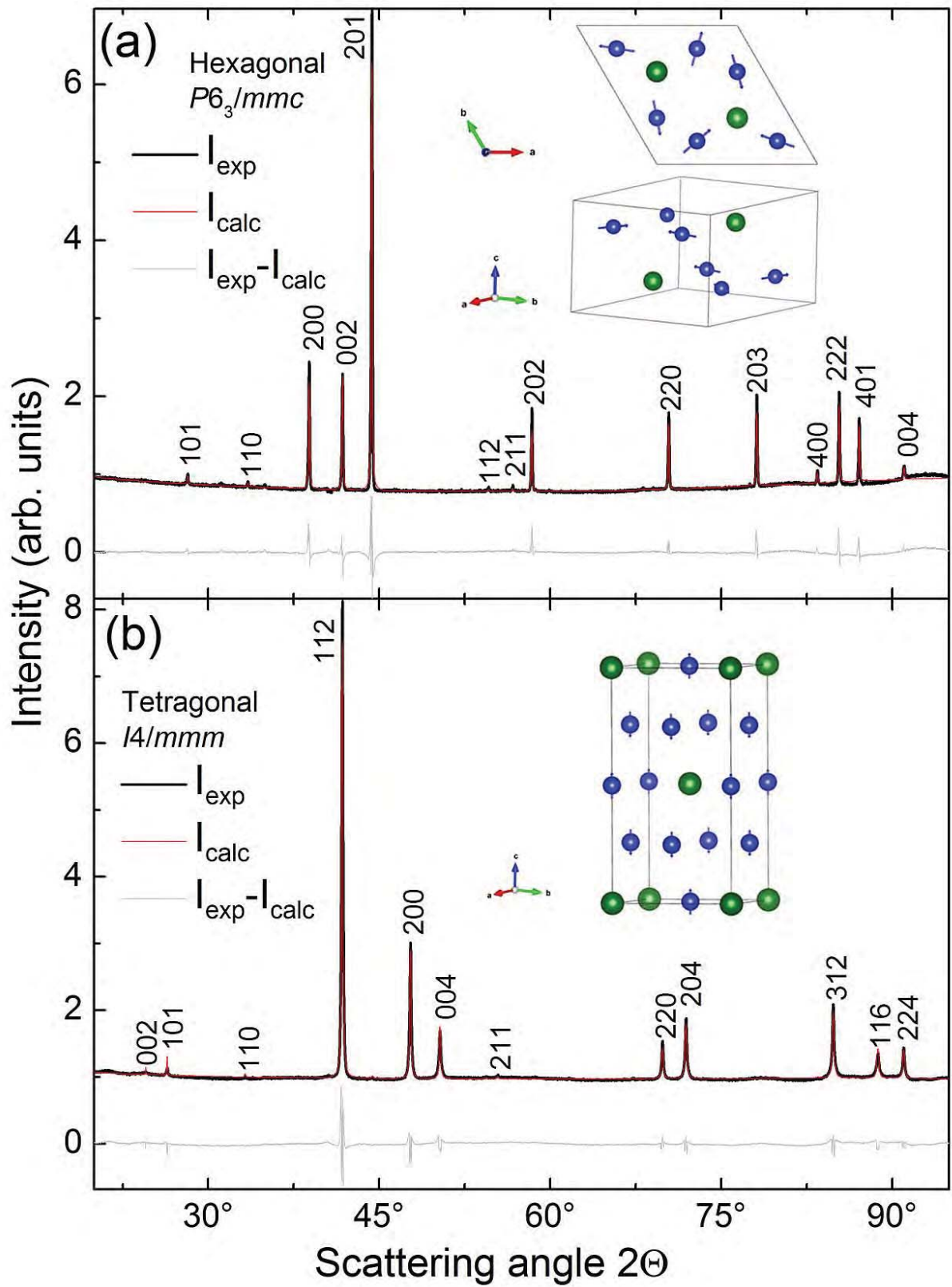

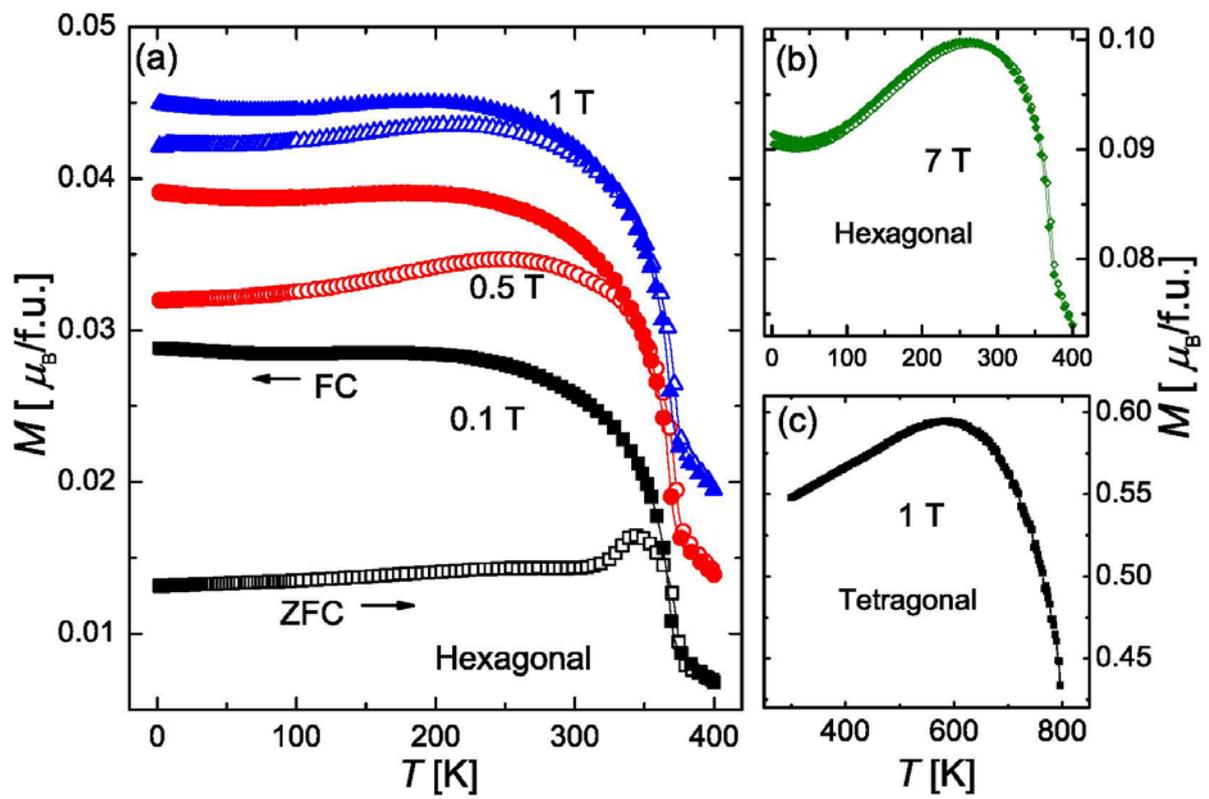

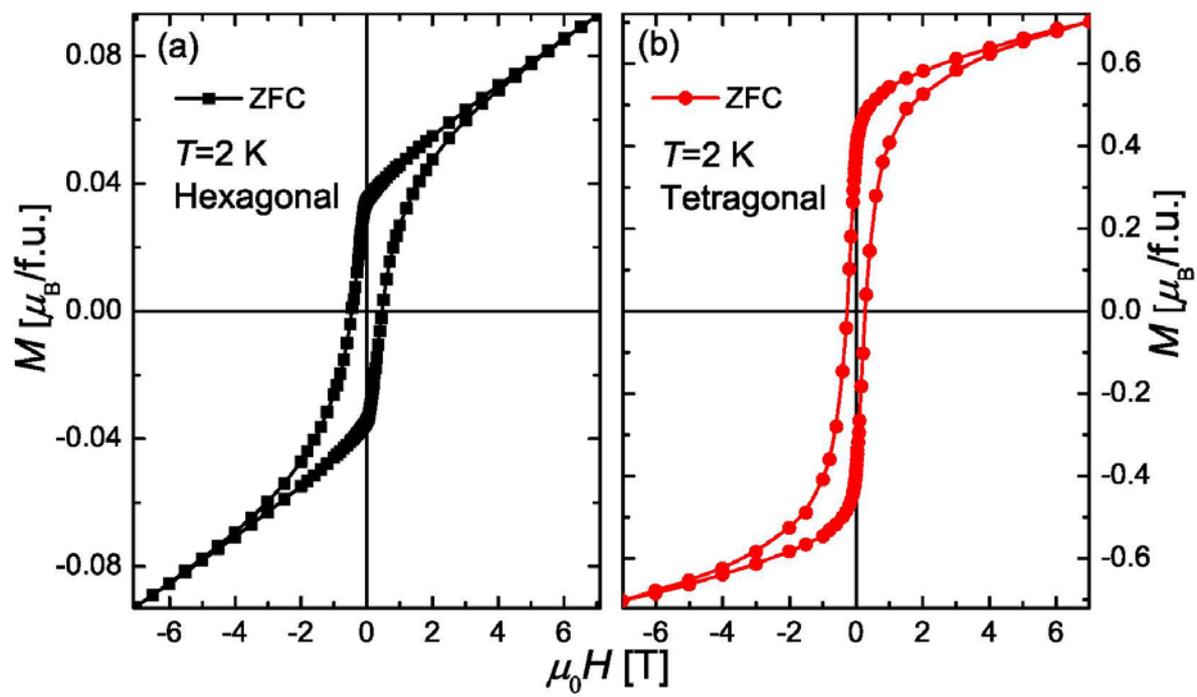

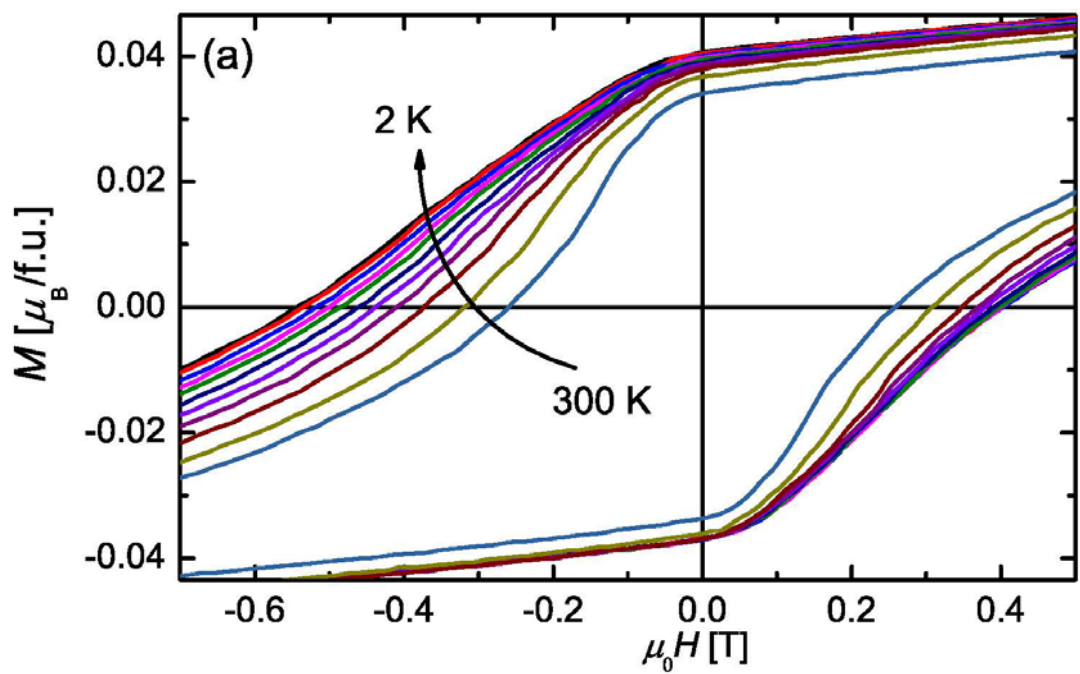
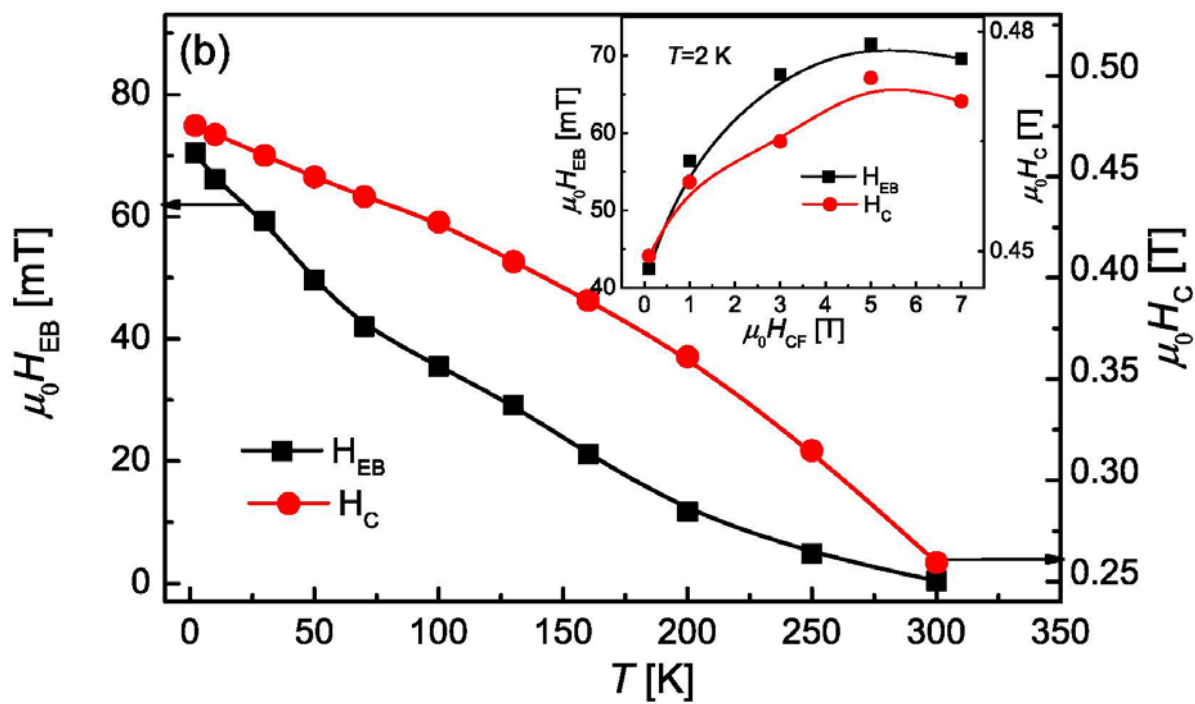